\documentclass[epj,nopacs]{svjour}
 \usepackage{mathrsfs}
\usepackage{amsmath, txfonts}
\usepackage{graphicx,bm,booktabs,bm}
\usepackage{longtable,lscape}
\usepackage{overpic,multirow,color}
\usepackage[colorlinks,
            citecolor=blue,
            anchorcolor=green,
            menucolor=orange,
            linkcolor=red,
            filecolor=red,
            runcolor=pink,
            urlcolor=blue,
            frenchlinks=red]{hyperref}

              \allowdisplaybreaks[4]

\graphicspath{{Figures/}} %

\begin{document}
\title{Interpretation of  $Y(4390)$ as an isoscalar partner of  $Z(4430)$ from  $D^*(2010)\bar{D}_1(2420)$ interaction}
\author{Jun He\inst{1}
\thanks{\emph{jun.he.1979@icloud.com (corresponding author)}}%
\and Dian-Yong Chen\inst{2}
\thanks{\emph{chendy@seu.edu.cn}}%
}                     
\institute{{Department of  Physics and Institute of Theoretical Physics, Nanjing Normal University,
Nanjing 210097, China}\and
{School of Physics, Southeast University, Nanjing 210094,  China}}

\date{Received: date / Revised version: date}
%
\abstract{
Invoked by the recent observation of  $Y(4390)$ at BESIII, which is about 40 MeV
below the $D^*(2010)\bar{D}_1(2420)$ threshold,  we investigate  possible bound and
resonance states from the  $D^*(2010)\bar{D}_1(2420)$  interaction
with the one-boson-exchange model in a quasipotential
Bethe-Salpeter equation approach. A bound state with  quantum number
$0^-(1^{--})$ is produced at 4384 MeV from the  $D^*(2010)\bar{D}_1(2420)$
interaction, which can be related to  experimentally observed
$Y(4390)$. Another state with quantum
number $1^+(1^{+})$ is also produced at $4461+i39$ MeV from this interaction. Different from the $0^-(1^{--})$ state,  the $1^+(1^{+})$  state is a resonance state  above the
$D^*(2010)\bar{D}_1(2420)$ threshold. This resonance state can
be related to the first observed charged charmonium-like state $Z(4430)$, which  has a mass about 4475 MeV measured  above the threshold  as observed
at Belle and LHCb.  Our result suggests that  $Y(4390)$
is an isoscalar partner of the  $Z(4430)$  as a hadronic-molecular state from  the $D^*(2010)\bar{D}_1(2420)$
interaction.
} 
\maketitle
\section{Introduction}\label{sec1}

A recent measurement of the cross section of $e^+e^-\to \pi^+\pi^-
h_c$ at center-of-mass energies from 3.896 up to 4.600 GeV suggested a
new resonance structure near the $D^*(2010)\bar{D}_1(2420)$ [thereafter
we denote it as $D^*\bar{D}_1$] threshold, $Y(4390)$, which has a mass of
$4391.5^{+6.3}_{-6.8}\pm 1.0$ MeV  and a width of
$139.5^{+16.2}_{-20.6}\pm0.6$ MeV~\cite{BESIII:2016adj}. After  observation of 
$Y(4390)$, a few interpretations of its internal structure were
proposed, such as a $3^3D_1$ charmonium state  in the conventional quark model
~\cite{Anwar:2016mxo}. A QCD sum rule calculation favors an assignment of
the $Y(4390)$ as a $D\bar{D}_1$ molecular state
~\cite{Wang:2016wwe}.  However, the $D\bar{D}_1$  threshold is much lower than  $Y(4390)$.
In fact, in the literature,  $Y(4260)$,
which is about 130 MeV lower than  $Y(4390)$, has been interpreted
as a $D\bar{D}_1$ molecular state~\cite{Chen:2016ejo,Cleven:2013mka}. Considering the $D^*$ meson is
about 140 MeV heavier than the $D$ meson, it is reasonable to discuss
an assignment of  $Y(4390)$ as a $D^*\bar{D}_1$ molecular state.

In the history of  study of  exotic states, the $D^*\bar{D}_1$ molecular state has been applied to interpret the first observed charged charmonium-like state near 4.43 GeV  with a mass of
$4433\pm4{\rm(stat)}\pm2{\rm(syst)}$ MeV and a width of
$45^{+18}_{-13}{\rm (stat)}^{+30}_{-13}{\rm(syst)}$ MeV reported by  Belle Collaboration~\cite{Choi:2007wga}. The mass measured by the Belle
Collaboration, about 4430 MeV, is close to the $D^*\bar{D}_1$ threshold, so it had ever been
popular to explain  $Z(4430)$ as an $S$-wave $D^*\bar{D}_1$ molecular
state with spin parity $J^P=0^-$~\cite{Close:2010wq,Liu:2008xz,Lee:2008tz}.
However,  a higher mass
of $4485^{+22+28}_{-22-11}$ MeV and a larger width of
$200^{+41+26}_{-46-35}$ MeV were reported by a new measurement  at Belle
Collaboration through a full amplitude analysis of $B^0\to\psi'
K^+\pi^-$ decay and a spin parity of $J^P=1^+$ was favored over other
hypotheses~\cite{Chilikin:2013tch}.   A new LHCb experiment
in the $B^0\to\psi'\pi^- K^+$ decay
confirmed the existence of the $1^+$ resonant structure $Z(4430)$  with
a mass of $4475\pm7^{+15}_{-25}$ MeV and a width of $172\pm13^{+37}_{-34}$
MeV ~\cite{Aaij:2014jqa}.

The new Belle and LHCb results support that the
spin parity of the $Z(4430)$ is $1^+$ instead of $0^-$ which was suggested by previous hadronic-molecular-state studies. If insisting on the interpretation of  $Z(4430)$ as a  $D^*\bar{D}_1$ molecular state, one  should go beyond  S wave, at least to P wave,  to reproduce experimentally observed  positive parity. Besides, the new measured mass of the $Z(4430)$ is higher than the $D^*\bar{D}_1$ threshold, which suggests that the $Z(4430)$ can not be a bound state.   To explain the new observation of  $Z(4430)$, Barnes $et\ al.$  suggested that the $Z(4430)$ is either
a $D^*\bar{D}_1$ state dominated by long-range $\pi$ exchange, or a
$D\bar{D}^*(1S, 2S)$ state with short-range
components~\cite{Barnes:2014csa}.  It has also been suggested that the
$Z(4430)$ may be from the $S$-wave $D\bar{D}'^*_1(2600)$ interaction, which has a threshold about 4470 MeV,
to avoid the difficulties mentioned above~\cite{Ma:2014zua}.

In Ref.~\cite{He:2014nxa}, the
$D^*\bar{D}_1$ and $D\bar{D}'^*(2600)$ interactions were  studied by solving the quasipotential
Bethe-Salpeter equation for vertex which is only valid for the bound state problem. It is found that  the $D\bar{D}'^*(2600)$ interaction is too weak to produce a bound state.  An isovector  bound state with quantum number $J^P=1^+$ can be produced from the $D^*\bar{D}_1$ interaction,  which corresponds to  $Z(4430)$.  Such a picture was confirmed by a lattice calculation where a state with $1^+(1^{+-})$ is also produced from the   $D^*\bar{D}_1$ interaction~\cite{Chen:2016lkl}. If  $Z(4430)$ is from the $D^*\bar{D}_1$ instead of  $D\bar{D}'^*(2600)$ interaction, the new observed $Z(4430)$ mass  at Belle and LHCb above the $D^*\bar{D}_1$ threshold suggests that  $Z(4430)$ should be a resonance state above the threshold instead of a bound state below the threshold.

In Refs.~\cite{He:2015cca,He:2015mja}, we develop a quasipotential Bethe-Salpeter equation for amplitude to study the resonance state above the threshold. With such formalism,  it is found that a state corresponding to  P wave should be taken as serious as these corresponding to  S wave~\cite{He:2016pfa}. Such an idea was applied to interpret the puzzling parities of  two LHCb hidden-charmed pentaquarks and  $Y(4274)$~\cite{He:2016pfa}. It was found that  a P-wave state is usually higher than an S-wave state because of  weaker interaction but is still hopefully to be observed. If we turn to the case of   $Y(4390)$ and $Z(4430)$, it is very natural to assign this two states as an S-wave   $D^*\bar{D}_1$ bound state and a P-wave  $D^*\bar{D}_1$ resonance state, respectively.  Hence, in this work we will investigate the $D^*\bar{D}_1$ interaction with the Bethe-Salpeter equation for the amplitude to study the possibility of interpreting the $Y(4390)$ and $Z(4430)$ as  hadronic molecular-states form the $D^*\bar{D}_1$  interaction.

In the next section, the formalism adopted in the current work is presented. The interaction potential is constructed with an effective Lagrangian and the quasipotential Bethe-salpeter equation will be introduced briefly. The numerical results are given in Section \ref{sec3}. A brief summary  is given in the last section.

\section{Formalism}\label{sec2}

In the energy region of  $Y(4390)$ and $Z(4430)$, besides the  $D^*\bar{D}_1$  threshold, there are  other three
thresholds of channels, $D^*$ $\bar{D}'_1(2430)$,
$D$$\bar{D}'^*(2600)$, and ${D}^*$$\bar{D}'^*(2550)$.  The large width of  $D'_1(2430)$, $\Gamma=384^{+130}_{-110}$ MeV~\cite{Agashe:2014kda}, which
means a very short lifetime, makes it difficult to bind  the $D^*$ meson and itself
together to form a state with a width of about $170$ MeV. The
 ${D}^*\bar{D}'^*(2550)$ interaction has also been related to the $Z(4430)$ in the literature. However, its threshold is about 100 MeV higher than
the $Z(4430)$ mass.  The calculation in Ref.~\cite{He:2014nxa} suggested that the $D\bar{D}'^*_1(2600)$ interaction and its coupling to the $D^*\bar{D}_1$ interaction are very weak. Hence, in this work, we only consider the $D^*\bar{D}_1$ interaction.

\begin{figure}[h!]\begin{center}
\includegraphics[bb=180 680 410 770,clip,scale=0.8]{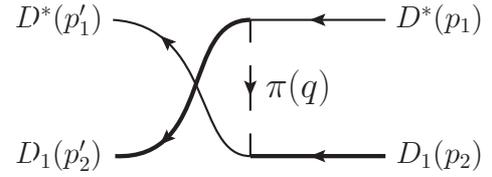}
\end{center}\caption{ The cross diagram  of the $D^*\bar{D}_1$ interaction by the $\pi$ exchange. \label{Fig: V}}
\end{figure}

For a loosely bound system, long-range interaction by the $\pi$ exchange should be more important than short-range interaction by exchanges of heavier mesons. The $Z(4430)$ locates higher than the $D^*\bar{D}_1$ threshold, so in the current work it will be seen as a resonance state where the interaction is even weaker than a loosely bound state. Hence, the dominance of the $\pi$ exchange is well satisfied in the case of  $Z_c(4430)$. However, in the case of  $Y(4390)$, the exchanges by heavier mesons may be involved because a binding energy about 40 MeV is found.  For the heavier pseudoscalar mesons which can  easily be introduced as $\pi$ meson in the frame of this work, their contribution is obviously much smaller than the $\pi$ meson because their mass is much heavier than the $\pi$ meson and the coupling constants are the same as those of the $\pi$ meson. An explicit calculation suggested the medium-range $\sigma$ exchange is very small partly due to its larger mass~\cite{He:2014nxa}. For the vector-meson exchange, it is difficult to determine the coupling constants involved with the existent information in the literature. And it is beyond the scope of this work to calculate such coupling constants. Furthermore, all vector mesons have much larger mass than $\pi$ meson, which also leads to a suppression effect on their contributions as in the case of the $\sigma$ meson exchange. Hence, in the current work, we do not include heavier-meson exchanges to avoid more not-well-determined coupling constants being introduced in the calculation with an assumption that the contributions form the heavier-meson exchanges are suppressed by the heavier mass as the $\sigma$ meson exchange. The direct diagram of the $\pi$ exchange  was also found negligible  compared with cross diagram by the $\pi$ exchange in an explicit calculation~\cite{He:2014nxa}. Hence, in this work, we will only consider  the cross diagram  of the $D^*\bar{D}_1$ interaction by $\pi$ exchange as shown in Fig.~\ref{Fig: V}.

The explicit flavor structures for isovectors ($T$) or isoscalars ($S$) $|D^*\bar{D}_1\rangle$ are \cite{Liu:2008xz}
\begin{eqnarray}
|D^*\bar{D}_1\rangle^+_T&=&\frac{1}{\sqrt{2}}\big(|D^{*+}\bar{D}^0_1\rangle+c|D^+_1\bar{D}^{*0}\rangle\big),\nonumber\\
|D^*\bar{D}_1\rangle^-_T&=&\frac{1}{\sqrt{2}}\big(|D^{*-}{D}^0_1\rangle+c|D^-_1{D}^{*0}\rangle\big),\nonumber\\
|D^*\bar{D}_1\rangle^0_T&=&\frac{1}{2}\Big[\big(|D^{*+}D^-_1\rangle-|D^{*0}\bar{D}^0_1\rangle\big)\nonumber\\
&+&c\big(|D^+_1D^{*-}\rangle-|D^0_1\bar{D}^{*0}\rangle\big)\Big],\nonumber\\
|D^*\bar{D}_1\rangle^0_S&=&\frac{1}{2}\Big[\big(|D^{*+}D^-_1\rangle+|D^{*0}\bar{D}^0_1\rangle\big)\nonumber\\
&+&c\big(|D^+_1D^{*-}\rangle+|D^0_1\bar{D}^{*0}\rangle\big)\Big],\label{flavor structure}
\end{eqnarray}
\normalsize
where $c=\pm$ corresponds to $C$-parity $C=\mp$. For the isovector state,  $c$ is related to the $G$-parity.

The involved effective Lagrangians describing the interaction between 
a light pseudoscalar meson $\mathbb{P}$ and heavy flavor mesons can be  constructed with the help of the
chiral  and heavy quark symmetries \cite{Isola:2003fh,Casalbuoni:1996pg},
\begin{align}\label{eq:lag-p-exch}
  \mathcal{L}_{D_1D^{*}\mathbb{P}}   &=
  i\sqrt{\frac{2}{3}}\frac{h'}{ f_\pi}
  \sqrt{m_{D_1}m_{D^{*}}}
 \nonumber\\
  &\cdot  \Big\{[-\frac{1}{4m_{D_1}m_{D^{*}}}D_{1b}^\alpha
  \overleftrightarrow{\partial}^\rho
  \overleftrightarrow{\partial}^\lambda
  D^{*\dag}_{\alpha a}\partial_\rho\partial_\lambda \mathbb{P}_{ba}
  \nonumber\\
  &-D_{1b}^\alpha  D^{*\dag}_{\alpha a}\partial^\rho\partial_\rho
  \mathbb{P}_{ba}+3D_{1b}^\alpha
  D^{*\dag\beta}_a\partial_\alpha\partial_\beta
  \mathbb{P}_{ba}]\nonumber\\
  &-[-\frac{1}{4m_{D_1}m_{D{*}}}D^{*\dag}_{\alpha a}
  \overleftrightarrow{\partial}^\rho
  \overleftrightarrow{\partial}^\lambda D_{1b}^\alpha
  \cdot\partial_\rho\partial_\lambda \mathbb{P}_{ab}
\nonumber\\
  &-
D^{*\dag}_{\alpha
a}D_{1b}^\alpha\partial^\rho\partial_\rho\mathbb{P}_{ab}
+3D^{*\dag\beta}_aD_{1b}^\alpha
\partial_\alpha\partial_\beta \mathbb{P}_{ab}]\Big\}.
\end{align}
With the above Lagrangians, we can obtain the potential for the cross diagram by  the $\pi$ exchange,
\begin{align}
i{\cal V}_{\lambda'_1\lambda'_2,\lambda_1\lambda_2}
&=f_I\frac{2}{3}\frac{h'^2m_{D_1}m_{D^*}}{
f^2_\pi(q^2-m_\pi^2)} ~\epsilon_{D_1,\lambda'_2}^{\dag\mu
}\epsilon^{\nu}_{D^*,\lambda_1}\epsilon_{D_1,\lambda_2}^{\rho}
 \epsilon^{\dag\sigma}_{D^*,\lambda'_1}
\nonumber\\&\cdot\left\{
\left[q^2-\frac{(p'^2_2-p_1^2)^2}{4m_{D_1}m_{D^*}}\right]g_{\mu\nu}-3q_{\mu}
q_{\nu}\right\}\nonumber\\
&\cdot
\left\{\left[q^2-\frac{(p^2_2-p'^2_1)^2}{4m_{D_1}m_{D^*}}\right]g_{\rho\sigma}-3q_{\rho}
q_{\sigma}\right\},\label{Eq: potential}
 \end{align}
where $p^{(')}_{1,2}$ and $\lambda^{(')}_{1,2}$ are  the initial (final) momentum and the helicity for
constituent 1 or 2.  And the flavor factor $f_I=-c/2$ and $3c/2$ for $I=1$ and $0$, respectively.
With  available
experimental information, Casalbuoni et al.
extracted $h'=0.55$
GeV$^{-1}$ from the old  data of decay
width $\Gamma_{tot}(D_1(2420))\approx 6$ MeV~\cite{Casalbuoni:1996pg}.  Compared with the new
suggested value of the decay width in  PDG, $25\pm6$ MeV~\cite{Agashe:2014kda}, a  value of  1.1 GeV$^{-1}$ can be obtained for the coupling constant $h'$. In this work, we will adopt this new value of $h'$ in the calculation.  The adoption of such value of $h'$ does not affect the  analysis above as regards the relative magnitude of  the contributions from different interaction channels and different exchanges.

The scattering amplitude of the $D^*\bar{D}_1$ interaction can be obtained by solving Bethe-Salpeter equation with the above potential. The Bethe-Salpeter equation is usually reduced to three-dimensional equation with a quasipotential approximation. To avoid the unphysical singularity from  the OBE interaction below the threshold, the off-shellness  of two constituent hadrons should be kept.  Here we adopt the most economic treatment, that is, the covariant spectator theory~\cite{Gross:1991pm,VanOrden:1995eg,He:2013oma},  which was explained explicitly in the appendices of Ref.~\cite{He:2015mja} and applied to a study of  the   $X(3250)$, the $Z(3900)$ and the LHCb pentaquarks and its strange partners~\cite{He:2012zd,He:2014nya,He:2015cea,He:2017aps}. In such a treatment,  we put the heavier constituent, $D_1$ meson here,  on shell~\cite{Gross:1999pd,He:2015yva}. Then the partial-wave Bethe-Salpeter equation with fixed
spin parity $J^P$ reads ~\cite{He:2015mja}
\begin{align}
i{\cal M}^{J^P}_{\lambda'_1\lambda'_2,\lambda_1\lambda_2}({\rm p}',{\rm p})
&=i{\cal V}^{J^P}_{\lambda'_1\lambda'_2,\lambda_1\lambda_2}({\rm p}',{\rm
p})+\sum_{\lambda''_1\lambda''_2\ge0}\int\frac{{\rm
p}''^2d{\rm p}''}{(2\pi)^3}\nonumber\\
&\cdot
i{\cal V}^{J^P}_{\lambda'_1\lambda'_2,\lambda''_1\lambda''_2}({\rm p}',{\rm p}'')
G_0({\rm p}'')i{\cal M}^{J^P}_{\lambda''_1\lambda''_2,\lambda_1\lambda_2}({\rm p}'',{\rm
p}),\quad\quad \label{Eq: BS_PWA}
\end{align}
Written down in the center-of-mass frame where $P=(W,{\bm 0})$, the
reduced propagator  is
\begin{align}
	G_0&=\frac{\delta^+(k''^{~2}_2-m_2^{2})}{k''^{~2}_1-m_1^{2}}
	\nonumber\\&=\frac{\delta^+(k''^{0}_2-E_2({\bm p}''))}{2E_2({\bm p''})[(W-E_2({\bm
p}''))^2-E_1^{2}({\bm p}'')]},
\end{align}
where the momentum of $D^*$ meson  $k''_1=(k''^{0}_1,-{\bm
p}'')=(W-E_2({\rm p}''),-{\bm p}'')$ and the momentum of the $D_1$ meson $k''_2=(k_2^{0},{\bm
p}'')=(E_2({\rm p}''),{\bm p}'')$ with $E_{1,2}({\rm p}'')=\sqrt{
M_{1,2}^{~2}+\rm p''^2}$. Here and hereafter we will adopt  a definition ${\rm p}=|{\bm p}|$.
The potential kernel ${\cal V}_{\lambda'_1\lambda'_2\lambda_1\lambda_2}$ obtained in  previous section,  the partial-wave potential with fixed spin parity $J^P$ can be calculated   as
\begin{eqnarray}
i{\cal V}_{\lambda'_1\lambda'_2\lambda_1\lambda_2}^{J^P}({\rm p}',{\rm p})
&=&2\pi\int d\cos\theta
~[d^{J}_{\lambda_1-\lambda_2\lambda'_1-\lambda'_2}(\theta)
i{\cal V}_{\lambda'_1\lambda'_2\lambda_1\lambda_2}({\bm p}',{\bm p})\nonumber\\
&+&\eta d^{J}_{\lambda_2-\lambda_1\lambda'_1-\lambda'_2}(\theta)
i{\cal V}_{\lambda'_1\lambda'_2-\lambda_1-\lambda_2}({\bm p}',{\bm p})],
\end{eqnarray}
where $\eta=PP_1P_2(-1)^{J-J_1-J_2}$ with $P_{(1,2)}$ and $J_{(1,2)}$ being the parity and spin of  constituent 1 or 2. Here without loss of  generality  the initial and final relative momenta can be chosen as ${\bm p}=(0,0,{\rm p})$  and ${\bm p}'=({\rm p}'\sin\theta,0,{\rm p}'\cos\theta)$, and the $d^J_{\lambda\lambda'}(\theta)$ is the Wigner d-matrix.

To guarantee the convergence of the integral in Eq.~(\ref{Eq: BS_PWA}), a regularization should be introduced. In this work we will introduce an exponential regularization by
a replacement of the propagator as
 \begin{eqnarray}
 G_0({\rm p})\to G_0({\rm
 p})\left[e^{-(k''^2_1-m_1^2)^2/\Lambda^4}\right]^2,\label{Eq: FFG}
 \end{eqnarray}
where $k''_1$ and $m_1$ are the momentum and mass of the lighter one of two constituent mesons. We would like to recall that the exponential  factor $e^{-(k''^2_2-m_2^2)^2/\Lambda^4}$ for particle 2  vanishes, which is only because the particle 2 is put on shell in the quasipotential approximation adopted in the current work. With such treatment, the contributions at large momentum ${\rm p}''$ will be suppressed heavily at the energies higher than 2 GeV as shown in Fig. 1 of Ref.~\cite{He:2016pfa}, and convergence of the integral is guaranteed. By multiplying the exponential factor on both sides of the Eq.~(\ref{Eq: BS_PWA}), it is easy to found that  the exponential factor can  also be seen as a form factor to reflect the off-shell effect of particle 1 in a form of $e^{-(k^2-m^2)^2/\Lambda^4}$. It is also the reason why a square of the exponential factor is introduced in Eq.~(\ref{Eq: FFG}).
The interested reader is referred to Ref.~\cite{He:2015mja} for
further information as regards  the regularization. A sharp cutoff of the momentum of  $p''$ at certain value $p''^{max}$, namely cutoff regularization,  is also often adopted in the literature~\cite{Oller:1998hw}. The exponential regularization can be seen as a soft version of the cutoff regularization. A comparison of the exponential regularization  and the  cutoff regularization as adopted in the chiral unitary approach~\cite{Oller:1998hw} was made in Ref.~\cite{He:2017aps} and it was found that the different treatments do not affect the conclusion. Because the current
treatment  guarantees the convergence of the integration, we do
not introduce the form factor for the exchanged meson, which is
redundant and its effect can be absorbed into  variation of the cutoffs $\Lambda$ as discussed
in Ref.~\cite{Lu:2016nlp}.

The integral equation ~(\ref{Eq: BS_PWA}) can be solved by  discretizing the momenta ${\rm p}$,
${\rm p}'$, and ${\rm p}''$ by the Gauss quadrature with a weight $w({\rm
p}_i)$. After such treatment, the integral equation can be transformed to a matrix equation  ~\cite{He:2015mja}
\begin{eqnarray}
{M}_{ik}
&=&{V}_{ik}+\sum_{j=0}^N{ V}_{ij}G_j{M}_{jk}.\label{Eq: matrix}
\end{eqnarray}
The propagator $G$ is a diagonal matrix with
\begin{eqnarray}
	G_{j>0}&=&\frac{w({\rm p}''_j){\rm p}''^2_j}{(2\pi)^3}G_0({\rm
	p}''_j), \nonumber\\
G_{j=0}&=&-\frac{i{\rm p}''_o}{32\pi^2 W}+\sum_j
\left[\frac{w({\rm p}_j)}{(2\pi)^3}\frac{ {\rm p}''^2_o}
{2W{({\rm p}''^2_j-{\rm p}''^2_o)}}\right],
\end{eqnarray}
with on-shell momentum
\begin{eqnarray}{\rm p}''_o=\frac{1}{2W}\sqrt{[W^2-(M_1+M_2)^2][W^2-(M_1-M_2)^2]}.\label{Eq: mometum onshell}
\end{eqnarray}

The scattering amplitude $M$ can be solved as  $M=(1-{ V} G)^{-1}V$.
Obviously,  the  pole of  scattering amplitude we wanted can be found at $|1-VG|=0$ after  analytic continuation total energy $W$
into the complex plane as $z$. In the current work, the pole is searched by scanning  the value of $|1-V(z)G(z)|$ by variation of real and imaginary parts of $z$, $Re(z)$ and $Im(z)$, in
complex plane to find  position of $z$ with  $|1-V(z)G(z)|=0$.

\section{The numerical results}\label{sec3}

\begin{figure}[h!]
\includegraphics[bb=55 54 410 300,clip,scale=0.7]{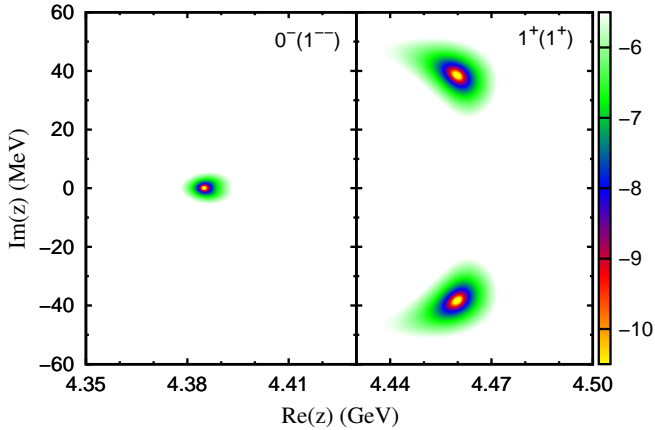}
\caption{ The $\log|1-V(z)G(z)|$ for the $D^*\bar{D}_1$ interaction . The results for the bound state with $0^-(1^{--})$ (left panel) and the resonance state with $1^+(1^{+})$ (right panel) are drawn to the same scale. \label{Fig: toy model}}
\end{figure}

With potential in Eq. (\ref{Eq: potential}),
the pole from the scattering amplitude can be found at $|1-V(z)G(z)|=0$ at complex plane by a continuation of the real center-of-mass energy $W$ to a complex $z$.  In this work,  only free parameter is the regularization cutoff $\Lambda$. By  varying  the cutoff, we try to found a  bound state with $0^-(1^{--})$ and a resonance state with $1^+(1^{+})$ which correspond to  $Y(4390)$ and  $Z(4430)$, respectively,  with the same cutoff.  In Fig.~\ref{Fig: toy model},  the  $\log |1-V(z)G(z)|$ is plotted  with variations of  Re($z$) and Im($z$). It is found that with cutoff $\Lambda=1.4$ GeV two states expected can be produced from the $D^*\bar{D}_1$ interaction.

Under the $D^*\bar{D}_1$ threshold,  a bound state with quantum number $0^-(1^{--})$  can be found at $z=4384$ MeV, which can be  obviously related to  $Y(4390)$ with a mass of 4391 MeV observed at BESIII. Since only the  $D^*\bar{D}_1$ interaction is considered in this work, no width is produced and the pole is at  real axis.  This state has a negative parity, so can be produced from the $D^*\bar{D}_1$ interaction in S wave.
For the state with  $1^+(1^{+})$, the P wave should be introduced to produce its positive parity.  As discussed in Ref.~\cite{He:2016pfa}, the P-wave interaction is  usually weaker  than the S-wave interaction. Furthermore,  for the $D^*\bar{D}_1$ interaction considered in this work, the flavor factor for the isoscalar sector is three times larger than that for the isovector sector, which makes the isovector interaction weaker.  Hence, one can expect that the $1^+(1^+)$ state is considerably higher than the  $0^-(1^{--})$ state. The result  in Fig.~\ref{Fig: toy model} confirms  such surmise. The expected $1^+(1^{+})$ state is found at $z=4461+i39$ which is much higher than  the $0^-(1^{--})$ state,   even above the $D^*\bar{D}_1$ threshold.  Obviously, this pole can be related to the charged charmonium-like state $Z(4430)$ whose  mass  is about 4475 MeV as suggested by the new LHCb experiment. Though only one-channel is included in this work, the resonance state carries a width as suggested by the scattering theory.

The above results show that the experimentally observed $Y(4390)$ and $Z(4430)$ can be reproduced from the  $D^*\bar{D}_1$ interaction with the same cutoff $\Lambda=1.4$ GeV.  In the rest part of this section, we will study whether there exist other possible states produced from this interaction.  Here, we only consider the  $D^*\bar{D}_1$ interaction with spin parties $0^\pm$, $1^\pm$, $2^{\pm}$, and $3^{-}$. Other partial waves are not considered because their spin parities cannot  be constructed with S and P waves.   Because a coupled-channel effect is not included in this work, we allow the regularization cutoff to deviate from the value above, 1.4 GeV,  by 0.5 GeV, i.e. from 0.9 to 1.9 GeV.  Only poles in an energy range  $4.35<Re(z)<4.50$ GeV are searched for in the calculation. The isovector states from the  $D^*\bar{D}_1$ interaction with typical cutoffs are listed in
Table~\ref{Tab: bound state1}.
\renewcommand\tabcolsep{0.25cm}
\renewcommand{\arraystretch}{1.5}
\begin{table}[h!]
\begin{center}
\caption{The  bound states from the $D^*\bar{D}_1$  interaction with typical cutoffs $\Lambda$.
The cutoff $\Lambda$ and  energy $W$ are in units of GeV, and MeV, respectively. \label{Tab:
bound state1}
\label{diagrams}}
	\begin{tabular}{cl|cl|clc}\bottomrule[1.5pt]
\multicolumn{2}{c|}{$1^+(0^{-})$}  &
\multicolumn{2}{c|}{$1^+(1^{+})$}   &  \multicolumn{2}{c}{$1^+(2^{-})$}\\\hline
 $\Lambda$ &  $W$   &  $\Lambda$ &  $W$  &  $\Lambda$ &  $W$\\\hline
 0.90   & 4471+i28   & 1.20 & 4492+i52  & 1.60  & 4454+i61  \\
 1.10   & 4459+i22   & 1.40  & 4461+i39 & 1.70  & 4429      \\
 1.30   & 4429       & 1.60  & 4431     & 1.80  & 4070      \\
 1.60   & 4368       & 1.80  & 4355     & 1.90  &  4358     \\  \hline
\toprule[1.5pt]
\end{tabular}
\end{center}

\end{table}

In the isovector sector, besides the $1^+(1^+)$ state corresponding to 
$Z(4430)$, there exist other  two possible states with
$1^+(0^-)$ and $1^+(2^-)$ produced
from the  $D^*\bar{D}_1$ interaction. With the decrease of the
regularization cutoff, the interaction gradually weaken.
As a result, the poles of  these states will run to
and cross the threshold at certain cutoff, and then the bound state
becomes a resonance state. If we fix the cutoff at 1.4 GeV as in the
case of reproducing  $Y(4390)$ and $Y(4430)$,  $1^+(0^-)$ is a
bound state around 4.4 GeV.  $1^+(2^-)$ is a resonance state much
higher  $Z(4430)$. A dependence of the results on the cutoff can be found in  Table 1, which is  from  neglecting of the coupled-channel effect and other approximations adopted in our approaches.  It is also the reason why we will vary the cutoff in the calculation, that is, the effects  of the approximations  can be absorbed into the variation of the cutoff.

The results of the isoscalar sector is listed in Table~\ref{Tab: bound
state2}.
Nine  states with $0^+(0^{\pm+})$, $0^\pm
(1^{+\pm})$, $0^\pm (1^{-\pm})$, $0^\pm(2^{+\pm})$ and $0^+(2^{-+})$
are produced in  this sector, which  are much more than three
states in the
isoscalar sector. It is reasonable because the flavor factor for the
isoscalar sector is three times larger than that for the isovector
sector, which means stronger interaction in this sector.
Generally speaking, the spin-negative states are more binding
than the positive states which reflects the P-wave interaction is
usually weaker than the S-wave state. In our calculation, more than one state is found in some cases, which can be seen
as excited state. As in the study of the hydrogen energy level  and the hadron spectrum in the constituent quark model, it is natural to find radial excited states besides the ground state. For the  $Y(4390)$ and  $Z(4430)$ which we focused on in this work, only one state was found in a considerable large range of the $Re(z)$,  and it is not so meaningful to present the results of excited states for other states even which ground state has not yet been observed in the experiment. So, in Tables~\ref{Tab: bound state1} and \ref{Tab: bound
state2}, only the results of the ground state are presented.

\renewcommand\tabcolsep{0.25cm}
\renewcommand{\arraystretch}{1.5}
\begin{table}[h!]
\begin{center}
\caption{The  bound states from the $D^*\bar{D}_1$  interaction with typical cutoffs $\Lambda$.
The cutoff $\Lambda$ and  energy $W$ are in units of GeV, and MeV, respectively. \label{Tab:
bound state2}
\label{diagrams}}
	\begin{tabular}{cl|cl|clc}\bottomrule[1.5pt]
\multicolumn{2}{c|}{$0^+(0^{++})$}&\multicolumn{2}{c|}{$0^+(0^{-+})$}   &  \multicolumn{2}{c}{$0^+(1^{++})$}\\\hline
 $\Lambda$ &  $W$   &  $\Lambda$ &  $W$  &  $\Lambda$ &  $W$\\\hline
 1.20   & 4449+i60   & 0.90 & 4200  & 0.90  & 4446+i18  \\
 1.40   & 4479+i66   & 1.00  & 4405 & 1.10  & 4422      \\
 1.60   & 4456+i72   & 1.10  & 4382 & 1.20  & 4401      \\
 1.70   & 4432+i78   & 1.20  & 4352 & 1.30  & 4368      \\
 \hline\hline
\multicolumn{2}{c|}{$0^-(1^{+-})$}  &\multicolumn{2}{c|}{$0^+(1^{-+})$}   &  \multicolumn{2}{c}{$0^-(1^{--})$}\\\hline
 $\Lambda$ &  $W$   &  $\Lambda$ &  $W$  &  $\Lambda$ &  $W$\\\hline
 1.40  & 4425 &  1.50  & 4428 & 1.15  & 4431   \\
 1.45  & 4413  & 1.60  & 4419 & 1.25  & 4425   \\
 1.50  & 4395 &  1.70  & 4392 & 1.35  & 4404   \\
 1.55  & 4374  & 1.80  & 4353 & 1.45  & 4365   \\
  \hline\hline
\multicolumn{2}{c|}{$0^+(2^{++})$}  &\multicolumn{2}{c|}{$0^-(2^{+-})$}   &  \multicolumn{2}{c}{$0^+(2^{-+})$}\\\hline
 $\Lambda$ &  $W$   &  $\Lambda$ &  $W$  &  $\Lambda$ &  $W$\\\hline
 1.10  & 4441+i5 & 1.20  & 4448+i24  & 1.00  & 4468+i47 \\
 1.30  & 4426    & 1.30  & 4431      & 1.10  & 4428     \\
 1.40  & 4395    & 1.40  & 4410      & 1.20  & 4419     \\
 1.45  & 4374    & 1.50  &  4371     & 1.40  & 4365     \\
\toprule[1.5pt]
\end{tabular}
\end{center}

\end{table}

\section{Summary}\label{sec4}

In this work, the  $D^*\bar{D}_1$ interaction is investigated in a
quasipotential Bethe-Salpeter equation approach, and  bound and
resonance states are searched for to interpret  $Y(4390)$ observed
recently at BESIII and the first observed charged charmonium-like state
$Z(4430)$.  A bound state with $0^-(1^{--})$ at 4384 GeV and a
resonance state with $1^+(1^{+})$ at $4461+i39$ MeV are produced from  the
$D^*\bar{D}_1$ interaction which can be related to  $Y(4390)$ and
 $Z(4430)$, respectively.  Hence,  $Y(4390)$ is an isoscalar
partner of  $Z(4430)$ and a partner of  $Y(4260)$ by replacing the $D$ meson by the $D^*$ meson in the hadronic-molecular state picture.

\vskip 10pt

\noindent {\bf Acknowledgement} This project is supported by the National Natural Science
Foundation of China (Grants No. 11675228  and No. 11375240), the Major State Basic Research Development
Program in China under grant 2014CB845405, and the Fundamental Research Funds for the Central Universities.

%


\end{document}